\documentclass[submission,copyright,creativecommons]{eptcs}

\providecommand{\event}{CompMod 2011}

\usepackage{cite}
\usepackage[pdftex]{graphicx}
\DeclareGraphicsExtensions{.pdf,.jpeg,.png}
\usepackage{amsmath,amsfonts,amssymb}
\usepackage{amsthm}
\usepackage[tight,footnotesize]{subfigure}
\usepackage{url}
\usepackage{color}
\providecommand{\firstpage}{1}

\theoremstyle{plain}
\newtheorem{theorem}{Theorem}[section]

\theoremstyle{definition}
\newtheorem{definition}{Definition}[section]

\theoremstyle{remark}

\newcommand{\vr}[1]{\mathbf{#1}}
\newcommand{\defeq}{\stackrel{def}{=}}
\newcommand{\bbR}{\mathbb{R}}

\newcommand{\calT}{\mathfrak{T}}
\newcommand{\calA}{\mathfrak{A}}
\newcommand{\calD}{\mathfrak{D}}
\newcommand{\sCCP}{\textbf{sCCP}}

\newcommand{\TC}{\mathcal{C}}
\newcommand{\TD}{\mathcal{D}}
\newcommand{\TS}{\mathcal{S}}
\newcommand{\init}[1]{init_{#1}}
\newcommand{\exit}[1]{\mathtt{q^1_{#1}}}
\newcommand{\enter}[1]{\mathtt{q^2_{#1}}}
\newcommand{\stoich}[1]{\mathtt{s}_{#1}}
\newcommand{\ratef}[1]{\mathtt{f}_{#1}}
\newcommand{\rate}[1]{\mathtt{\lambda}_{#1}}
\newcommand{\cmode}[1]{\mathtt{q}_{#1}}
\newcommand{\priority}[1]{\mathtt{w}_{#1}}
\newcommand{\guard}[1]{\mathtt{G}_{#1}}
\newcommand{\reset}[1]{\mathtt{R}_{#1}}

\begin{document}

\title{Programmable models of growth and mutation\\ of cancer-cell populations}

\author{Luca Bortolussi
\institute{ Dept. of Mathematics and Informatics, University of
Trieste, Italy.\\
\email{luca@dmi.units.it}} \and Alberto Policriti \institute{ Dept.
of Mathematics and Informatics,
University of Udine, Italy.\\
Institute of Applied Genomics, Udine, Italy.\\
\email{alberto.policriti@uniud.it}} }
\def\titlerunning{Programmable models of cancer cells}
\def\authorrunning{L. Bortolussi \& A. Policriti}

\maketitle

\begin{abstract}
In this paper we propose a systematic approach to construct mathematical models describing populations of cancer-cells at different stages of disease development. The methodology we propose is based on stochastic Concurrent Constraint Programming, a flexible stochastic modelling language. The methodology is tested on (and partially motivated by) the study of prostate cancer. In particular, we prove how our method is suitable to systematically reconstruct different mathematical models of prostate cancer growth --together with interactions with different kinds of hormone therapy-- at different levels of refinement.

\noindent \textbf{Keywords:} Computational Systems Biology, Stochastic Process Algebras, Tumour Growth Modelling, Prostate Tumour.

\end{abstract}


\section{Introduction} 
\label{sec:intro}

The (long term) goal of this work is the development of a general
framework for building programmable models to be used in the study
of cancer. The models we have in mind are, in fact, models of
populations of cells obeying specific growth and death laws and
incorporating different stages of tumour evolution.

At the architectural level, we envisage three key features:
\begin{itemize}
  \item \emph{programmability}, as the possibility to enter a high-level
  description of the full model in terms of networks of
  interacting agents;
  \item \emph{hybrid-ness}, as the possibility to carry out a complete description of the semantics of our
  networks of agents by a dynamical system with a controlled
  combination of both discrete and continuous evolution;
  \item \emph{stochasticity}, as the possibility of specifying the interaction
  mechanisms  obeying to given stochastic laws.
\end{itemize}

The above mentioned networks naturally implement the hybrid nature
of the different phases through which the development of tumoral
populations undergo. Moreover, they  allow us to  study both the
logical and quantitative features of a model, at a higher level of description
than classical models based on Ordinary Differential Equations
(ODE). As an example of such analyses, one can try to understand to
what extent the observed \textit{noise} depends on the structure of the internal/external
interactions defining the model, as opposed to parameter variation.
Finally, they allow  to tune-up the level of discreteness to be
used in the modelling activity.

Our concrete and motivating example is the study of prostate cancer,
with special emphasis on the predictiveness ability of models
including external interactions (in the form of medicament
dispensation)~\cite{BIO:CANCER:Brawer:2006:HormonalTherapyProstate,SB:Aihara:2010:prostateRoyalSoc}.
For this reason we begin illustrating  the various models of prostate cancer-cell evolution under different policies of chemical castration (see
Section~\ref{sec:background}
and~\cite{SB:Aihara:2010:prostateRoyalSoc}). 

Our modelling approach to describe prostate cancer is based on
stochastic Concurrent Constraint Programming
(\sCCP,~\cite{SB:Bortolussi:2008:BiomodelingSCCP:Journal}) a
modelling language belonging to the class of stochastic Process
Algebras (SPA) applied to Systems Biology \cite{SB:HillstonCiocchetta:2008:ProcessAlgebraSysBio}.
SPA have been applied in modelling many aspects of biological systems, including tumour growth \cite{SB:Lecca:2011:TumorShrinkage, SB:Mazza:2009:TumorGrowthP}. 

The objectives of this work are twofold: First we want to describe prostate cancer models and drug dispensation policies in \sCCP. \sCCP\ models of prostate tumour are presented in
Section~\ref{sec:model}. Secondly, we want to clarify if  noise observed in experimental data could be
explained as a structural feature of the model. A first experimental
study in this sense is presented in Section~\ref{sec:results}.

\sCCP\ is particularly suitable for the proposed task because its
semantics is naturally stochastic, it is easily programmable and
extensible, and it also has a general semantics in terms of
stochastic hybrid automata~\cite{SB:Bortolussi:2009:CompMod} (see
Sections~\ref{sec:sCCP},~\ref{sec:TDSHA}, and~\ref{sec:HybridSCCP}).
However, in order to properly use \sCCP\ for cancer modelling, we had
to enlarge the set of primitives that can be used to describe agent
behaviour. In particular, we had to introduce a mechanism to
describe actions triggered by conditions on model time and to give
the agents the ability of changing their environment (i.e. system
variables) according to random laws. These extensions turned out to
be very simple to introduce in the hybrid semantics framework of
\sCCP\ (cf. Section~\ref{sec:extensionSCCP}).

%
%


\section{Background}
\label{sec:background}

Below we briefly discuss the two main objects of our work in this paper: prostate cancer modelling  techniques and
 \sCCP, the modelling language that we will use to build our
agent network in the rest of the paper.

\subsection{Prostate cancer modelling}
\label{sec:prostateModel}

Prostate is a gland of the male reproductive system responsible for
the production of seminal fluid. Prostate tumour is a very common
(age related) disease, consisting mainly in the development of a
mass of tumoral cells whose growth is correlated with (dependent
from) the presence of androgen hormones (e.g.
testosterone)~\cite{BIO:CANCER:Brawer:2006:HormonalTherapyProstate}.
Most effective therapies consist in androgen deprivation
(castration) by surgical or chemical
means~\cite{BIO:CANCER:Brawer:2006:HormonalTherapyProstate}. Under
such therapies, one observes an initial fast decrease of tumoral
masses which, however, after a variable interval of time, undergo a
relapse phase. This is caused by the emergence of a line of androgen
independent tumoral cells, resistant to androgen
deprivation~\cite{BIO:CANCER:Brawer:2006:HormonalTherapyProstate,BIO:CANCER:Abrahamsson:2010:IASprostateCancerReview}.

The above biological behaviours have been modelled using
phenomenological
approaches~\cite{SB:Jackson:2004:prostateModelCAS,SB:Aihara:2008:prostateIAS,SB:Aihara:2010:prostateRoyalSoc}
based on ODEs. The variables of the model record the number of
androgen dependent/independent cells and are equipped with growth
and death laws expressing their time evolution. The spatial structure of the tumour is ignored, only the cell number is recorded. The observable of
the model is the serum Prostate Specific
Antigen~\cite{BIO:CANCER:Rao:2008:PSAdiscovery} (PSA)---a bio-marker
whose concentration in serum is strictly related to the size of the
tumour mass, that is the number tumoral cells.

\begin{figure}[!t]

$$
\begin{array}{cc}
\begin{array}{c}
   \begin{array}{rcl}
     \frac{dx}{dt} & = & G_x(x,z) - D_x(x,z) - M_{xy}(x,z)\\
     \frac{dy}{dt} & = & G_y(y,z) - D_y(y,z) + M_{xy}(x,z)\\
     \frac{dz}{dt} & = & P_z - D_z(z)\\
  \end{array}\\
  \\
  P_z = 0\ \ \ D_z(z) = \frac{z}{\tau}
\end{array}
&
\begin{array}{l}
G_x(x,z) = \alpha_x\left(k_1+(1-k_1)\frac{z}{z+k_2}\right)x\\
D_x(x,z) = \beta_x\left(k_3+(1-k_3)\frac{z}{z+k_4}\right)x\\
G_y(y,z) = \alpha_y\left(1-d\frac{z}{z0}\right)y\\
D_y(y,z) = \beta_y y\\
M_{xy}(x,z) = m_1\left(1-\frac{z}{z_0}\right)\\
\end{array}

\end{array}$$

  \caption{{\footnotesize ODE model of prostate tumor growth under Continuous Androgen Suppression,
  taken from~\cite{SB:Aihara:2010:prostateRoyalSoc}. $P$ stands for \emph{Production}, $G$ for \emph{Growth},  $D$ for \emph{Death}, and $M$ for \emph{Mutation}. In order to control $P_z$ the parameter $\alpha_z$ is set to $0$ to
  represent the effect of androgen deprivation therapy. In absence
  of castration, such a rate would be $P_z = \frac{z_0}{\tau}$.
  Parameters are set according to~\cite{SB:Aihara:2010:prostateRoyalSoc}:
  $\alpha_x = 0.0204$, $\alpha_y = 0.0242$, $\beta_x = 0.0076$, $\beta_y = 0.0168$, $k_1 = 0$, $k_2 = 2$,
  $k_3 = 8$, $k_4 = 0.5$, $m_1 = 5\cdot 10^{-5}$,  $z_0 = 20$, $d = 1$, $\tau = 62.5$.
We agree that $\tau$ controls the speed of $z$ dynamics,  $z_0$ is
the stationary value of androgen hormone in absence of chemical
castration, and $d\in [0,1]$ controls the   effect of androgen
hormone in the growth rate of AI cells (0 means that growth is
independent from the hormone, 1 means that the growth is inhibited
by the hormone). Notice how both the growth and death terms for
AD-cells $x$ depend non-linearly on the amount of hormone $z$
through a sigmoid-shaped function $\frac{z}{z+k}$} }\label{tab:ODE}
\end{figure}

More specifically, a model can be defined by the system of differential
equations of Figure~\ref{tab:ODE}, which corresponds to the initial model used in~\cite{SB:Jackson:2004:prostateModelCAS} and~\cite{SB:Aihara:2010:prostateRoyalSoc}. In the equations, we
have three variables: $x$ describes the population of Androgen Dependent (AD) cancer
cells, $y$ describes the population of Androgen Independent (AI) cancer cells, and $z$ is
the concentration of androgen hormone. The concentration of PSA is computed simply as $x+y$. This model is able to capture
the relapse phase of the tumour due to androgen independent
cells growth~\cite{SB:Jackson:2004:prostateModelCAS,SB:Aihara:2008:prostateIAS}.

A (clever) clinical approach tackling the problem of relapse after a
long interval of chemical castration, is  the so-called Intermittent Androgen
  Suppression (IAS) policy, see~\cite{BIO:CANCER:Abrahamsson:2010:IASprostateCancerReview}.
The effectiveness of the IAS therapy is based on the fact that the underlying mechanism of the disease seems to involve a
%
%
%
  competition between androgen dependent and androgen independent
 cells. This implies that a complete reduction of the number of
  AD cells removes any obstacle to AI cells growth. Hence, a more
effective strategy consists in maintaining a certain number of AD
cells to inhibit AI cells growth. The overall effect of an
intermittent androgen deprivation strategy is the delay---possibly
for a very long time---of the development of an androgen independent
tumour.

Given the quantities involved and the presence of an external
input---the medicament dispensation strategy---to keep into account
as a time-controlled phase change, the mathematical modelling
machinery naturally evolved into a hybrid model. To be precise, the
system can switch between states
representing normal and androgen deprivation
modes~\cite{SB:Aihara:2008:prostateIAS,SB:Aihara:2010:prostateRoyalSoc}.
These modes can be conveniently represented by a boolean variable
$u$, where $u = 1$ describes the drug dispensation phase. The
equations are obtained from the set of Figure~\ref{tab:ODE} by
replacing $P_z$ in the equation for the androgen hormone $z$ by the
$u$-dependent function $P_z(u) = \frac{z_0(1-u)}{\tau}$.
Note that $P_z(1) = 0$, as in the previous model. A
further evolution of the model consisted in the introduction of
noise in the equation, thereby moving to a set of
\textit{stochastic} differential equations. This move was performed
as an attempt to capture small fluctuations observed in the PSA
data.

Our first motivation was to attempt to clarify whether the noise
introduced in the model of Aihara et al.
\cite{SB:Aihara:2010:prostateRoyalSoc} is external or
internal. In the former case, the explanation would call into play
additional (noise) sources not related with growth and death laws of
AD/AI cells. In the latter case, the noise could be explained in
terms of fluctuations of such transitions, when considered as
stochastic. Our approach is to begin designing a discrete and stochastic
version of the above model, in terms of Continuous Time Markov
Chains
(CTMC)~\cite{STOC:Norris:1997:MarkovChains,SB:Gillespie:1977:gillespieAlgorithm}.
The construction will be given in Section \ref{sec:model} and the
experimental results will be given in Section \ref{sec:results}.
Below  we briefly describe our agent's language, that will act both as
an intermediate layer in the translation to CTMC and as a
computational counterpart of the model based on (stochastic)
differential equations.

\subsection{Stochastic Concurrent Constraint Programming}
\label{sec:sCCP}

We briefly introduce now (a simplified version of) stochastic
Concurrent Constraint Programming (\sCCP), sketching the basic
notions needed in the rest of the paper. More details on the
language can be found
in~\cite{SB:Bortolussi:2008:BiomodelingSCCP:Journal}. \sCCP\ has two
basic ingredients: \emph{agents} and \emph{constraints}. Agents are
the main actors, interacting by asynchronously exchanging
information, in form of \emph{constraints}, through the
\emph{constraint store}. \sCCP\ has been mainly applied as a
modelling language for biological
systems~\cite{SB:Bortolussi:2008:BiomodelingSCCP:Journal}, using the
constraint store to describe the state of the system, e.g.
numerosity of molecular species. These quantities are described by a
set of variables that can change value during computation, called
\emph{stream
variables}~\cite{SB:Bortolussi:2008:BiomodelingSCCP:Journal}. At
least for modelling simple biological scenarios, one needs very
simple constraints, basically comparing and assigning new values to
stream variables.

\begin{definition}\label{def:sCCP}
An \sCCP\ program is a tuple $\calA = (A,\calD,\vr{X},init(\vr{X}))$,
where
\begin{enumerate}
\item The \emph{initial network of agents} $A$
and the \emph{set of definitions} $\calD$ are given by  the
following grammar:
$$\begin{array}{c}
A =  M~|~A\parallel A \ \ \ \
M = \pi.C~|~M+M\ \ \ \
\pi =  [G(\vr{X}) \rightarrow R(\vr{X},\vr{X'})]_{\lambda(\vr{X})}\\
\calD =  \emptyset~|~\calD \cup\calD~|~\{C\defeq M\} \\
\end{array}$$
\item $\vr{X}$ is the set of stream variables of the store (with global scope), usually taking integer
values;
\item $init(\vr{X})$ is a predicate on $\vr{X}$ of the form $\vr{X}
= \vr{x_0}$, assigning an \emph{initial value} to store
variables.
\end{enumerate}
\end{definition}

In the previous definition, basic actions are \emph{guarded updates}
of (some of the) variables: $G(\vr{X})$ is a quantifier-free first
order formula whose atoms are inequality predicates on variables
$\vr{X}$ and $R(\vr{X},\vr{X'})$ is a predicate on $\vr{X},\vr{X'}$
of the form  $\vr{X'} = \vr{X} + \vr{k}$ ($\vr{X'}$ denotes
variables of $\vr{X}$ after the update)\footnote{The constraints
that can be used to update the constraint store are rather limited,
as they simply add a constant to some stream variables. This
restriction, however, allows to interpret \sCCP-actions as
continuous fluxes, a required condition to define the hybrid
semantics, see also Section~\ref{sec:HybridSCCP}.}, for some vector
$\vr{k}\in\bbR^n$. Each such action has a \emph{stochastic
duration}, specified by associating an exponentially distributed
random variable to actions, whose rate depends on the state of the
system through a function $\lambda$, with values
$\lambda(\vr{X})\in\bbR^+$. The semantics of \sCCP\ is given by a
Continuous Time Markov Chain~\cite{STOC:Norris:1997:MarkovChains},
deduced from the \textit{labelled transition
system} associated with an agent network  by the operational semantics \cite{SB:Bortolussi:2008:BiomodelingSCCP:Journal}.

All \sCCP\ agents $C\defeq M \in \calD$, according to
Definition~\ref{def:sCCP}, are \emph{sequential}, i.e. they do not
contain any occurrence of the parallel operator, whose usage is
restricted at the upper level of the network. \sCCP\ sequential
agents can be seen as automata synchronizing on store variables and
they can be conveniently represented as labelled graphs, called
\emph{Reduced Transition Systems} (RTS)
(see~\cite{PA:Bortolussi:2009:SCCPandODEjournal}). More precisely,
$RTS(C) = (S(C),E(C))$ is a multi-graph with vertices $S(C)$
corresponding to the different states of an agent and with edges
$e\in E(C)$ corresponding to actions performable by the agent, labelled by the
corresponding rate ($\lambda_e$), guard ($G_e$), and reset
($R_e$).
In the following, we also need the
notion of extended \sCCP\ program $\calA^+$, in which we introduce a
new  set  of variables $\vr{P} = \{P_C~|~C \in \calD\}$ taking integer values
and recording how many copies of agent $C\in\calD$ are in parallel
in the current \sCCP-network. Rates, guards, and resets are modified
to consistently treat   $\vr{P}$-variables. With this trick, we can
assume that in $\calA^+$ there is never more than one copy of the
same agent running in parallel at the same time. Further details on
these notions can be found
in~\cite{PA:Bortolussi:2009:SCCPandODEjournal}.

Examples of \sCCP-programs will be given below and throughout the
rest of the paper. We begin by observing that playing with the meaning
assigned to variables and transitions, models at different levels of
detail can be programmed in a multi-level environment. At the lowest
level, even a single cell can be represented by an agent, thereby
suggesting a programming style such as:
{\small
\begin{eqnarray*}
\mathtt{cell\_H}(M) &\mbox{:-}&  [M<\mu_0 \rightarrow M'=M+1]_{\lambda_h(M)}.\mathtt{cell\_H}(M)
                    +   [M=\mu_0 \rightarrow M'=M]_{\lambda_m}.\mathtt{cell\_T}(M) \\
\mathtt{cell\_T}(M) &\mbox{:-}& [M<\mu_1 \rightarrow M'=M+1]_{\lambda_t(M)}.\mathtt{cell\_T}(M)
                     + [M=\mu_1 \rightarrow M'=M]_{\lambda_d}.\mathbf{0} \\
\end{eqnarray*} }
In the previous piece of code, we are describing the mutation from
a cell in a healthy state ($\mathtt{cell\_H}$) to a cell in a
tumoral state ($\mathtt{cell\_T}$), as a consequence of accumulation
of (internal) genomic mutations. The number of mutations of a cell
is represented by a (local) variable $M$, which is increased by one
when a mutation event occurs. A cell will remain in a healthy state
as long as the total number of mutations is bounded by $\mu_0$
(first action). We assumed that mutations can happen at a rate
dependent on the internal state of the cell (number of mutations
$M$). When such a critical level is reached, the cell after some
time will become tumoral (second action). A tumoral cell, instead,
will survive while the number of genomic mutations remains below
$\mu_1$ (third action). When such a threshold is reached, the cell
will die (fourth action). A dead cell is represented by the null
agent $\mathbf{0}$, which is removed from the system.

The full program  would run in parallel a number of agents equal to
the number of cells modelled, each with its specific
mutation-counting variable. We remark that this very simple model,
not including events like cell duplication and cell apoptosis, has been introduced just for illustrative purposes.

The above (low) level  of modelling will not be used in the
following, even though along the lines of the above example the
competition mechanism between AD and AI cells could be explicitly
programmed. Instead, we will use variables counting the number of AD
and AI cells, and assign agents to interactions, like cell population
growth/death. This corresponds to ``programming the dynamics'' at a higher
level of abstraction. We will discuss this usage in the next
section.

%
%

Even if the standard semantics of an \sCCP-program is given in terms
of CTMC, \sCCP-programs have also different semantics, based on
ODE~\cite{PA:Bortolussi:2009:SCCPandODEjournal} and hybrid
automata~\cite{SB:Bortolussi:2009:CompMod}, providing an additional
degree of flexibility to the language. In particular, the hybrid
semantics is parametric with respect to the degree of continuity
introduced in the model, and it subsumes as special cases both the
CTMC and the ODE semantics. We will give now more details on this
semantics, exploiting it in Section~\ref{sec:extensionSCCP} to
smoothly extend \sCCP\ with additional primitives that we will need
for modelling tumour-cells.\footnote{A software tool to model and analyse \sCCP\ programs is under development. A preliminary version is available from the authors upon request.}

\subsection{Transition-Driven Stochastic Hybrid Automata}
\label{sec:TDSHA}

\emph{Transition-Driven Stochastic Hybrid Automata} (TDSHA) have
been introduced
in~\cite{SB:Bortolussi:2009:CompMod,PA:Bortolussi:2010:HybridDynamicsStochProg:TCS}
as a convenient formalism to associate a stochastic hybrid automaton
to an \sCCP\ program.
\\
The emphasis of TDSHA is on \emph{transitions} which, as always in
hybrid automata, can be either discrete (corresponding to jumps) or
continuous (representing flows acting on system's variables). More specifically, there are two kinds of discrete transitions:
\begin{enumerate}
\item \emph{instantaneous} or \emph{forced}: taking place as soon as their
guard becomes true, and
\item \emph{stochastic}: occurring after an
exponentially distributed delay.
\end{enumerate}

\begin{definition}\label{def:TSHS}
A Transition-Driven Stochastic Hybrid Automaton (TDSHA) is a tuple\\
$\calT = (Q,\vr{X},\TC,\TD,\TS,\init{})$, where:
\begin{itemize}
\item $Q$ is a finite set of \emph{control modes} and $\vr{X} = \{X_1,\ldots,X_n\}$ is a set of real valued
\emph{system's variables}\footnote{Notation: the time
  derivative of $X_j$ is denoted by $\dot{X_j}$, while the value
of $X_j$ after a change of mode is indicated by $X_j'$}.

\item $\TC$ is the set of \emph{continuous transitions or flows},
containing triples $\tau=(q,\vr{s},f)$, where: $q\in Q$ is a
mode, $\vr{s}$ is a real vector of size $|\vr{X}|$, and
$f:\bbR^n \rightarrow \bbR$ is a (Lipschitz) function. They are
indicated by $\cmode{\tau}$, $\stoich{\tau}$, and
$\ratef{\tau}$, respectively.

\item $\TD$ is the set of \emph{discrete or instantaneous transitions},
whose elements are tuples $\eta=(q_1,q_2,G,R,w)$, where: $q_1$
is the \emph{exit-mode},  $q_2$ is the \emph{enter-mode}, and
$w:\bbR^n\rightarrow \bbR^+$ is a weight function used to
resolve non-determinism between two or more active transitions.
Moreover, $G$ is a quantifier-free first-order formula with free
variables among $\vr{X}$ and $R$, which is a conjunction of atoms of
the form $X' = r(\vr{X},\vr{\mu})$, where
$r:\bbR^{n+m}\rightarrow \bbR$, is the reset function of $X$,
depending on system's variables $\vr{X}$ as well as  on a vector of
parameters $\vr{\mu}$. In particular, parameters in $\mu$ can be
either constants or random values. The elements of a tuple
$\eta$ are indicated by $\exit{\eta}$, $\enter{\eta}$,
$\priority{\eta}$, $\guard{\eta}$, and $\reset{\eta}$,
respectively.

\item $\TS$ is the set of \emph{stochastic transitions}, whose
elements are tuples $\eta = (q_1,q_2,G,R,\lambda)$, where $q_1$,
$q_2$, $G$, and $R$ are as for transitions in $\TD$, while
$\lambda:\bbR^n\rightarrow \bbR^+$ is a function giving the
state-dependent rate of the transition. Such function is
referred to by $\rate{\eta}$.

\item $\init{}$ is a pair $(q_0,\vr{X_0}) \in Q\times\bbR^n$,
identifying the \emph{initial state} of the system.
\end{itemize}
\end{definition}

The dynamics of TDSHA can be formally
defined~\cite{SB:Bortolussi:2009:CompMod} by associating with
them a Piecewise Deterministic Markov
Processes~\cite{STOC:Davis:1993:PDMP}. Intuitively, the TDSHA
dynamics is given by periods of continuous evolution,
interleaved by discrete jumps, as customary with hybrid systems.
\\
Within each mode $q\in Q$, the TDSHA evolves following the solution
of a set of Ordinary Differential Equations (ODE), constructed
combining the different continuous transitions active in $q$. Each
such $\tau \in \TC$ contributes to the ODE with a flow given by
$\ratef{\tau}(\vr{X})$, whose effect on each variable is described
by $\stoich{\tau}$. Hence, in mode $q\in Q$, ODE are given by
$\dot{\vr{X}} = \sum_{\tau\in\TC{}, \exit{\tau} = q}
\stoich{\tau}\ratef{\tau}(\vr{X})$.
\\
Two kinds of discrete jumps are possible: stochastic transitions
$\eta\in\TS$ are fired according to their rate, while
instantaneous transitions $\eta\in\TD$ are fired as soon as
their guard becomes true. In both cases, the state of the system
is reset according to the policy specified by $\reset{\eta}$.
Choice among several  active stochastic or instantaneous
transitions is resolved probabilistically according to their
rate or priority.
\\
We stress that TDSHA have another sources of stochasticity in
addition to priorities and rates: resets.
Resets function, in fact, can assign random values to variables,
through their dependency on random variables (via the parameter's
tuple $\mu$). This feature of TDSHA will be exploited in
Section~\ref{sec:extensionSCCP}.
\\
Furthermore, a notion of asynchronous TDSHA product $\calT =
\calT_1\otimes \calT_2$ can be defined in a simple way,
see~\cite{PA:Bortolussi:2010:HybridDynamicsStochProg:TCS}.
Essentially, the discrete state space  of the product automaton
is $Q_1\times Q_2$, while transitions from state $(q_1,q_2)$ are
all those issuing from $q_1$ \emph{or} $q_2$.

\subsection{Hybrid Semantics of sCCP}
\label{sec:HybridSCCP}
In this section we briefly recall the definition of the hybrid
semantics for \sCCP~\cite{SB:Bortolussi:2009:CompMod}, which is
given by associating a TDSHA to an \sCCP\ program. The mapping is
compositional: first single \sCCP\ components are converted to
TDSHAs, then one TDSHA is obtained as their product.

A TDSHA is, ultimately, an automaton and the discrete skeleton of the TDSHA associated to an \sCCP\ component $C$ is directly derived from the RTS --which is a labelled graph-- of $C$. The level of
discreteness/continuity will be a parameter, which is fixed by
partitioning RTS-edges $E(C)$ into discrete edges $E_d(C)$ and
continuous edges $E_c(C)$. The former ones will generate stochastic transitions and the latter ones
continuous transitions. The edge partition is encoded by a boolean
vector $\kappa$ of size $E(C)$, where $\kappa[e] = 1$ means that
$e\in E_c(C)$. The recipe for constructing the TDSHA
$\calT_{C,\kappa} = (Q_C,\vr{Y},\TC_C,\TD_C,\TS_C,\init{C})$ of
component $C$ of $\calA^+$ is sketched below.

\

\noindent\textbf{Discrete Modes}. Continuous edges in $E_c(C)$ can
short-circuit different RTS-states of $S(C)$. Hence, we collapse
together RTS-states connected by an (undirected) path of continuous
edges. Let $[C_i]_c$ be the set---equivalence class---of states containing $C_i\in S(C)$.
Then $Q_C = \{[C_i]_c~|~C_i\in S(C)\}$.
\\
\textbf{Variables}. The TDSHA uses exactly the variables $\vr{Y}$ of the
extension $\calA^+$:  system variables $\vr{X}$ and state variables
$\vr{P}$.
\\
\textbf{Continuous transitions}. Each edge in $e\in E_c(C)$ becomes
a transition in $(q,\vr{s},f)\in\TC_C$, where: $q = [C_i]_c$, with
$C_i$ the source state of $e$, $\vr{s}$ is the update vector
defining $R_e$ (i.e. $R_e$ equals $\vr{Y}'=\vr{Y} + \vr{s}$), and
$f(\vr{Y}) = \lambda_e(\vr{Y})$.
\\
\textbf{Stochastic transitions}. Each edge $e\in E_d(C)$ produces a
discrete stochastic transition $(q_1,q_2,G,R,\lambda)\in\TS_C$,
where $q_1$ is the equivalence class containing the source state $C_1$ of $e$, $q_2$ contains
the target state $C_2$ of $e$, $G\defeq G_e$, $\lambda \defeq
\lambda_e$, and $R\defeq R_e\wedge statevar\_res$. The second term
of $R$ is needed to correctly deal with classes of states of $S(C)$:
$statevar\_res \defeq (\bigwedge_{C\in q_1} P_{C}' = 0) \wedge
P_{C_2}'=1$.
\\
\textbf{Initial conditions}. $\init{C}$ is constructed by
considering the initial state of the \sCCP\ component in $\calA^+$.
\\
\textbf{Instantaneous transitions}. Instantaneous transitions are
not needed at this level of the mapping. They have been used
in~\cite{SB:Bortolussi:2009:CompMod} to describe dynamic partition
policies between discrete and continuous transitions, and they will
be used in the next section to give \sCCP\ new functionalities.

\

Once we have constructed the TDSHA for each parallel component of $\calA^+$ corresponding to the initial network $A = C_1\parallel\ldots\parallel
C_n$, we apply the (basically) standard product construction  introduced at the end of \ref{sec:TDSHA} to obtain
$\calT_{A,\kappa_1\oplus\ldots\oplus\kappa_n} =
\calT_{C_1,\kappa_1}\otimes\cdots\otimes\calT_{C_n,\kappa_n}$.

\paragraph{\textbf{Lattice of TDSHA}.} The reader can easily see that the previous definition is parametric
with respect to the degree of continuity introduced in the hybrid
automata. We can arrange the different TDSHA obtained by different
choices of $\kappa$ in a lattice, where the top element is the fully
continuous and deterministic TDSHA, while the bottom element is the
fully discrete and stochastic TDSHA. In particular, the fully
continuous TDSHA has one single mode, no discrete transitions, and
it corresponds to the set of ODE associated to an \sCCP\ program by
fluid-flow
approximation~\cite{PA:Bortolussi:2009:SCCPandODEjournal}. On the
other hand, the fully discrete TDSHA is the CTMC associated with an
\sCCP\ by the standard semantics.

\

We stress the fact that \sCCP\ is not a language designed to describe stochastic hybrid systems, but  it is rather a programming language to model networks of (stochastically) interacting agents. The hybrid semantics has been defined as a computationally efficient approximation of the stochastic one. Consequently, the \sCCP\ syntax has no way of specifying if an action has to be interpreted as discrete or continuous. This choice has to be performed at the RTS level. Practically, we are investigating the definition of general rules providing good automatic partitions (in terms of behaviour approximation and computational efficiency) and the external annotation of \sCCP\ actions by naming them, similarly to Bio-PEPA~\cite{SB:HillstonCiocchetta:2009:bioPEPA}. As a final remark, we point out that the asynchronous nature of \sCCP\ allows the definition of the hybrid semantics in a relatively simple compositional way.


\section{Extending sCCP: Events and Random Updates}
\label{sec:extensionSCCP}

In this section we will extend the \sCCP\ language by equipping it with additional primitives that  will be used, in our case study, to model prostate tumour dynamics. In particular, in order to model drug delivery policies, we need to describe SBML-like \cite{SB:SBMLwebsite,SB:Ciocchetta:2009:BioPEPAevents} 
events in \sCCP. Furthermore, in the experiments we will carry out in Section~\ref{sec:results}, we will also need to reset variables to random values, drawn according to given distributions. In the following, we will briefly introduce these extensions, both syntactically and semantically. More specifically, we will add new syntactic primitives to \sCCP, framing their semantics within the skeleton of TDSHA, extending the hybrid semantics discussed in the previous section.

\paragraph{\textbf{Events.}}
SMBL-like events describe instantaneous modifications of the system,
triggered by conditions on the system state or on the simulation
time.  Events allow us to model very easily external influences
on a system, such as, for instance, the hormone deprivation policies for
prostate cancer.
More precisely, following~\cite{SB:Ciocchetta:2009:BioPEPAevents},
an SBML-like event is a conditional expression of the form

\begin{center}
\texttt{if (trigger) then (assignments) with (delay)},\\
\end{center}
where \texttt{trigger} is a condition involving system variables and
the simulation time, \texttt{assignment} is an update of (some)
system variables, and \texttt{delay} is an optional time delay
between the activation of the trigger and the consequent update of
the variables.

We will first discuss how to describe in \sCCP\ events with trigger depending  system's  state only and with no delay. In order to do this, we simply have to extend \sCCP\ with instantaneous actions, i.e. actions taking no time to occur. Syntactically, this can be done by allowing rates to take value infinity, i.e. we have to allow basic actions of the form $[G\rightarrow R]_\infty$. Even if instantaneous transitions are quite standard in stochastic process algebras \cite{PA:Hermanns:2000:PAforPerformanceEvaluation,PA:Bernardo:1998:EMPAtcs,PA:Balbo:1995:ModelingGSPN}, we will take care to define their semantics within the TDSHA framework. In particular, we will map an action $ C = [G\rightarrow R]_\infty.C_1 + M$ to the instantaneous transition $([C]_c,[C_1]_c,G,R,w)$, where the priority $w$ is the constant function 1 (i.e. $w(\vr{X}) = 1$ for each $\vr{X}$).  Therefore, these \sCCP\ actions are always kept discrete while mapping \sCCP-programs to TDSHA. If we consider the TDSHA obtained by keeping all \sCCP\ transitions discrete, then the \sCCP\ semantics with instantaneous actions remains a CTMC, provided the \sCCP\ program cannot execute an infinite number of instantaneous actions in the same time instant (a well-known fact, cf.~\cite{PA:Balbo:1995:ModelingGSPN}), because non-determinism between two or more active instantaneous transitions of TDSHA is resolved probabilistically by the use of priorities.

Time controlled events, instead, are more difficult to express in stochastic process algebras, as they change the semantics of the language, from a CTMC to a \textit{semi}-Markov process~\cite{PA:Mode:2005:SemiMarkovProcesses}, i.e. a stochastic process in which transition times can be sampled according to general distributions. In particular, we are considering semi-Markov processes mixing exponential and deterministic transition times. Such processes can be simulated using priority queues, as customary done in discrete event simulation. The introduction of time controlled events in \sCCP, however, is done by simply adding a special (\textit{reserved)} variable, $Time$, to the language. The use of variable $Time$ is restricted to infinite-rate actions, like $[Time = 10 \rightarrow R]_\infty$, and it \emph{cannot} be reset.

From a semantic point of view, the value of $Time$ corresponds to
the simulation time of the model. This fact can be
built into the TDSHA framework: we add $Time$ to the TDSHAs variables
and we render its dynamics by adding a new automaton (sometimes  called
\textit{time-monitor}) in parallel with the TDSHA obtained from the \sCCP\
program. This automaton will have one single mode $q$, one variable
($Time$), initial state $(q,0)$, and one single continuous
transition, $(q,\vr{s},f)$---modeling the flow of time with rate
$f=1$ and stoichiometric vector $\vr{s} = \vr{1}$. As all actions
containing $Time$ have infinite rate, they will all become
instantaneous transitions in the TDSHA. Therefore, an action with
guard $Time  = 10$ will activate as soon as TDSHA variable $Time$,
which corresponds to the model time, will reach value 10. Of course, the
guards of infinite-rate actions can combine in complex ways
conditions on time and on system variables.
Notice that events with a delay $d$ between the activation of the trigger and the consequent assignment of system's variables can be easily modelled within \sCCP\ as a sequence of two instantaneous actions: $[trigger \rightarrow T' = Time + d]_\infty.[Time = T \rightarrow assignment]_\infty$. We can formally render the correctness of this construction in the following

\begin{theorem}\label{thSemiMarkov}
The TDSHA obtained from an \sCCP\ program with instantaneous transitions and time constraints, keeping all the non-instantaneous transitions stochastic and coupling it with a time-monitor, is stochastically equivalent to a semi-Markov process.
\end{theorem}

\paragraph{\textbf{Random Updates.}}
The capability of updating system variables with random values can be useful in many applications, and it will be exploited in this paper to model uncertainty and intrinsic variability in (kinetic) parameters. For instance, we can model an action whose rate is a random variable rather than a number, abstracting in this way an (unknown) underlying mechanism giving rise to such randomness.

Alternatively, random resets can be used to incorporate uncertainty about a parameter in the model, according to the Bayesian point of view \cite{SB:Wilkinson:2006:StochasticModellingSB}. In particular, we can select randomly the value of a parameter at the beginning of each simulation according to a given prior distribution (reflecting our knowledge on the parameter value). Then, the empirical distribution of the certain and the uncertain model are compared, assessing the sensitivity of the model with respect to the parameter.\footnote{An alternative way to incorporate lack of knowledge is to use imprecise probabilities, like in Imprecise Markov Chains \cite{StocSkulj2009ImpreciseMC}.}

To formally add random resets in \sCCP, we allow arbitrary reset functions in infinite-rate \sCCP\
actions $[G\rightarrow R]_\infty$. Therefore, the reset $R$ of such
an actions can now be a conjunction of atoms of the form $X' =
R(\vr{X},\vr{P},\vr{\mu})$, where $\mu$ is a vector of parameters,
which can be constants\textit{ or} random variables, like in the definition
of TDSHA. The mapping from \sCCP\ to TDSHA does not need to be
modified: the instantaneous TDSHA transition associated to an
infinite-rate action is the one described for events. We observe
that there is no real obstacle in using general resets also in
\sCCP\ actions with finite rate, but in this case these actions
cannot be approximated continuously in the hybrid semantics, but
need to be kept discrete.

We stress that the advantage of defining the semantics of these extensions in terms of TDSHA (instead of defining them at the operational semantics level, which is  possible at least for random resets) is that they apply to all semantics of \sCCP, including the CTMC and ODE based semantics, which are special instances of the TDSHA semantics.

Equipped with these extensions to \sCCP, we now focus on prostate tumour growth modelling. Our first goal is to build a discrete and stochastic (programmable) model, capable of rendering the prostate tumour-growth model(s) presented in Section~\ref{sec:prostateModel}. This approach will be discussed in the following section, while the effects of noise in this model will be discussed in Section~\ref{sec:results}.


\section{sCCP model of prostate cancer growth}
\label{sec:model}
In this section we describe how to go from a mathematical model like
the one of Section~\ref{sec:background} to \sCCP-programs,
illustrating the technique directly on the prostate tumour case.

The basic principles in our approach consist in identifying from the
differential equations the \sCCP\ \emph{variables} and the \sCCP\
\emph{interactions}. Variables will roughly correspond to variables
in the differential equation model. Interactions will be obtained by
disassembling the right-end sides of the differential equations,
identifying explicitly the different actions modifying the populations. Interactions will be described by agents of the
network.

For prostate tumour cancer,
following~\cite{SB:Aihara:2010:prostateRoyalSoc}, we use four
variables $X, Y, Z,$ and $V$ standing for the (numbers of) AD cells,
AI cells, androgen hormone molecules, and PSA molecules,
respectively.

It will be convenient to classify interactions into two classes:
cellular interactions (i.e. growth, death, mutation) and molecular
interactions (i.e. production, degradation). For prostate tumour
cancer we have five cellular interactions, corresponding to growth
and death for AI and AD cells, in addition to an interaction
modelling mutation of AD into AI cells.  As far as molecular
interactions are concerned, we consider four of them, that is
production and degradation of androgen hormone and PSA. The
differential equation model defines the PSA level as the sum of the
number of AI and AD cells. We have chosen, instead, to treat PSA
production and degradation explicitly, in order to give more
internal flexibility to the model.

\begin{figure}[!t]

{\footnotesize
\begin{tabular}{c}
  \begin{tabular}{cc}
  $\mathtt{growthAD}\ \mbox{:-}\ [X>0 \rightarrow X' = X +
     1]_{G_X(X,Z)}.\mathtt{growthAD}$ &
  $\mathtt{deathAD}\ \mbox{:-}\ [X>0 \rightarrow X' = X -
     1]_{D_X(X,Z)}.\mathtt{deathAD}$ \\
  $\mathtt{growthAI}\ \mbox{:-}\ [Y>0 \rightarrow Y' = Y +
     1]_{G_Y(Y,Z)}.\mathtt{growthAI}$ &
  $\mathtt{deathAI}\ \mbox{:-}\ [Y>0 \rightarrow Y' = Y -
     1]_{D_Y(Y,Z)}.\mathtt{deathAI}$\\
  $\mathtt{produceANDH}\ \mbox{:-}\ [true \rightarrow Z' = Z +
     1]_{P_Z}.\mathtt{produceANDH}$ &
  $\mathtt{degradeANDH}\ \mbox{:-}\ [Z>0 \rightarrow Z' = Z -
     1]_{D_Z(Z)}.\mathtt{degradeANDH}$\\
  $\mathtt{producePSA}\ \mbox{:-}\ [true \rightarrow V' = V +
     1]_{P_V(X,Y)}.\mathtt{producePSA}$ &
  $\mathtt{degradePSA}\ \mbox{:-}\ [V>0 \rightarrow V' = V -
     1]_{D_V(V)}.\mathtt{degradePSA}$
  \end{tabular}\\
  $\mathtt{mutateADtoAI}\ \mbox{:-}\ [X>0 \rightarrow X' = X -
    1\wedge Y'=Y+1]_{M_{XY}(X,Z)}.\mathtt{mutateADtoAI}$\\
  \begin{tabular}{ccc}
    $G_X(X,Z) = \alpha_x\left(k_1+(1-k_1)\frac{Z}{Z+k_2 \Omega_Z}\right)X$ &
    $D_Y(Y,Z) = \beta_y Y$ &
    $D_Z(Z) = \frac{Z}{\tau}$\\
    $D_X(X,Z) = \beta_x\left(k_3+(1-k_3)\frac{Z}{Z+k_4\Omega_Z}\right)X$ &
    $M_{XY}(X,Z) = m_1\left(1-\frac{Z}{z_0\Omega_Z}\right)$ &
    $P_V(X,Y) = \frac{\Omega_V(X+Y)}{N_0}$\\
    $G_Y(Y,Z) = \alpha_y\left(1-d\frac{Z}{z0\Omega_Z}\right)Y$&
    $P_Z = 0$ &
    $D_V(V) = V$
  \end{tabular}\\
\end{tabular} }

\caption{{\footnotesize List of agents of the \sCCP-program modelling
the prostate tumour growth. The initial configuration of the network
consists of all the agents above, running in parallel. Rates are
obtained from the ODE of Section~\ref{sec:background} by suitably
scaling parameters according to conversion factors, to be denoted
$\Omega_Z$, $\Omega_V$, and $N_0$. $\Omega_V$ is the conversion
factor between PSA concentration and PSA numerosity. It corresponds
to the approximate number of molecules in a nanogram per millilitre
of PSA. Analogously, $\Omega_Z$ is the conversion factor for $Z$,
i.e the number of molecules giving a concentration of a nano-mole
per litre. $N_0$, instead, is the reference number of cells, defined
as the number of cells that on average produce $1\frac{ng}{ml}$
units of PSA. The production and degradation rates of PSA are
defined so that the average stationary number of PSA molecules is
$\frac{\Omega_V(X+Y)}{N_0}$ corresponding to a concentration equal
to $x + y$, which is the PSA value computed in the ODE model of
Section~\ref{sec:background}. }}\label{tab:sccpProgram}
\end{figure}

In Figure~\ref{tab:sccpProgram} the reader can find the full network
of agents corresponding the equations presented in Section
\ref{sec:background}. In order to illustrate how agents interact, a
precise definition of rates must be given. Consider, for example,
the agent corresponding to growth of AD cells:
$$\mathtt{growthAD}\ \mbox{:-}\ [X>0 \rightarrow X' =
X + 1]_{G_X(X,Z)}.\mathtt{growthAD}$$
The agent represents a loop in
which the number of AD cells grows by one at each iteration. The
growth takes place only when the guard $X>0$ is satisfied, and it
has the effect of incrementing the value of $X$ by one unit. The
rate of the interaction is directly derived from the differential
equation models:
$$G_X(X,Z) = \alpha_x \left(k_1 + (1-k_1)\frac{Z}{Z+ k_2\Omega_Z} \right)X. $$
The parameter $\Omega_Z$ appearing in the above rate, is not found
in the differential equation model: it is a scaling
parameter necessary to convert the concentration $z$ of the hormones
into the molecular count $Z$. Other conversion factors are required
for the size of cell populations $X$ and $Y$ ($N_0$) and for
PSA concentration ($\Omega_V$), see caption of
Figure~\ref{tab:sccpProgram}. The significance of these conversion
factor in the dynamics of the entire model will be discussed in the
Section \ref{sec:results}.

\begin{figure}[!t]
  \begin{center}
  \subfigure[{\footnotesize ODE vs Stochastic, CAS policy}]{\includegraphics[width
=0.48\textwidth]{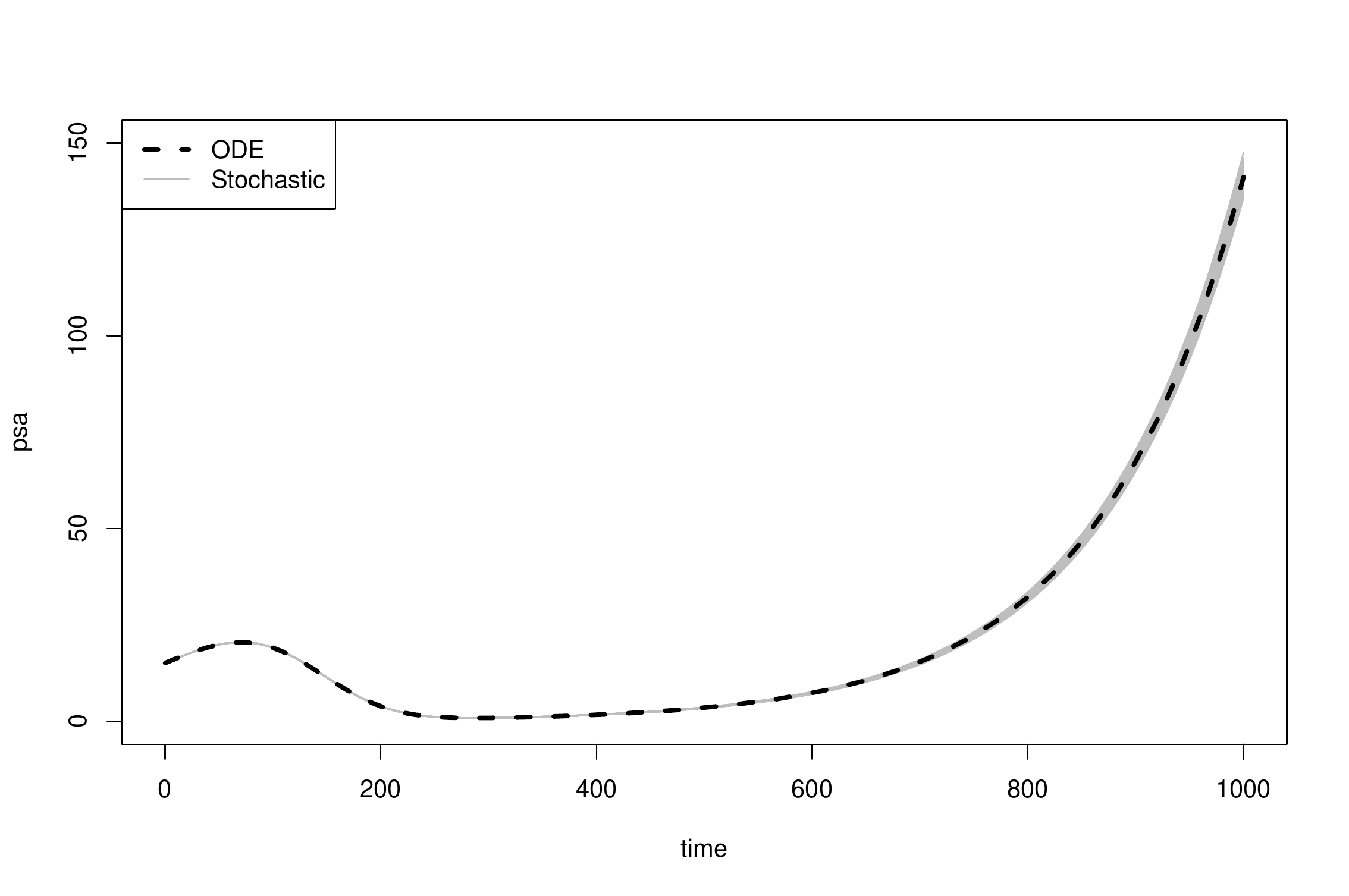} \label{figCAS} } 
\subfigure[{\footnotesize ODE vs Stochastic, IAS policy}]{\includegraphics[width
=0.48\textwidth]{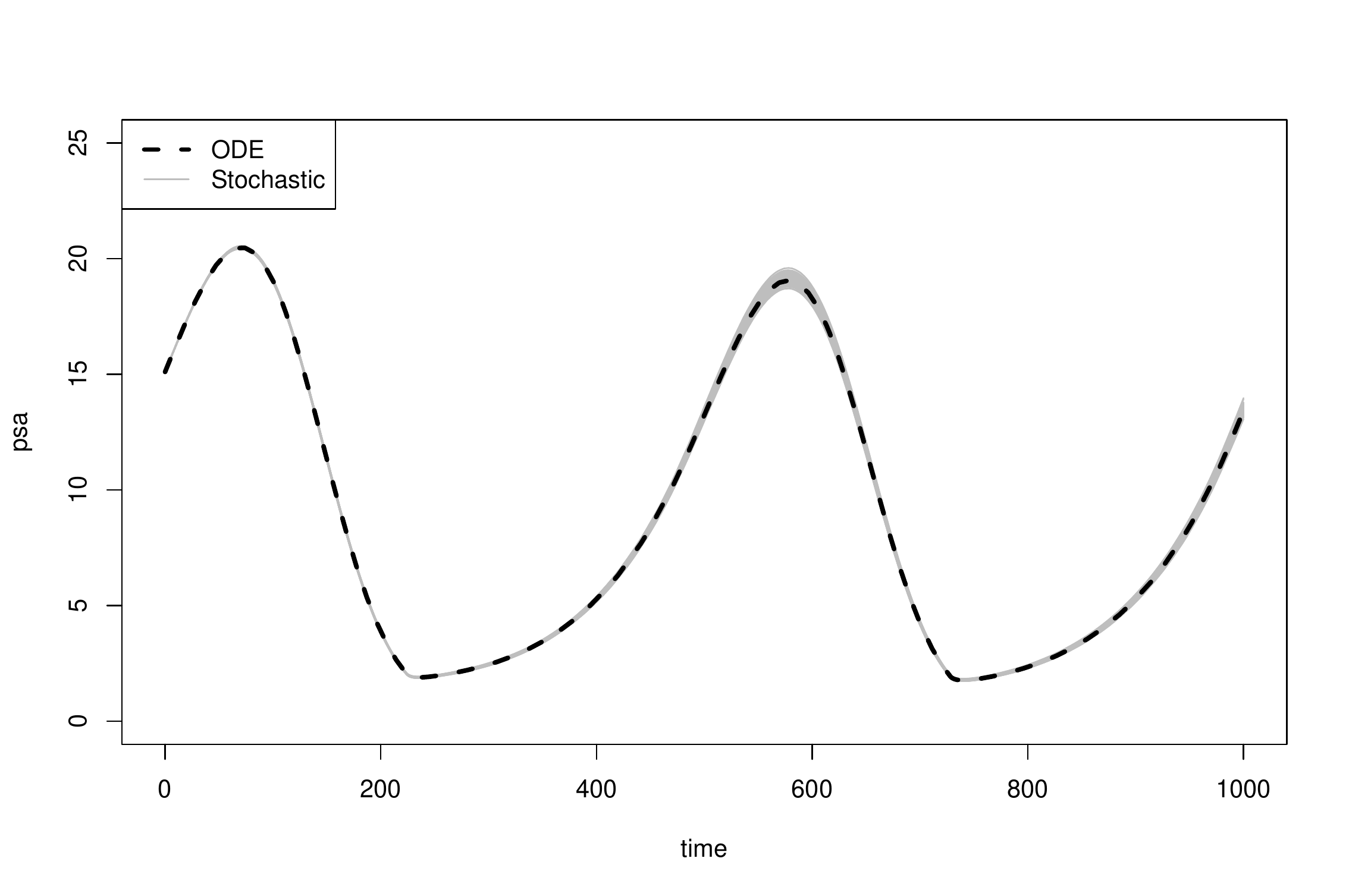} \label{figIAS}}

\subfigure[{\footnotesize CAS stochastic, $N_0$ varying}]{\includegraphics[width
=0.48\textwidth]{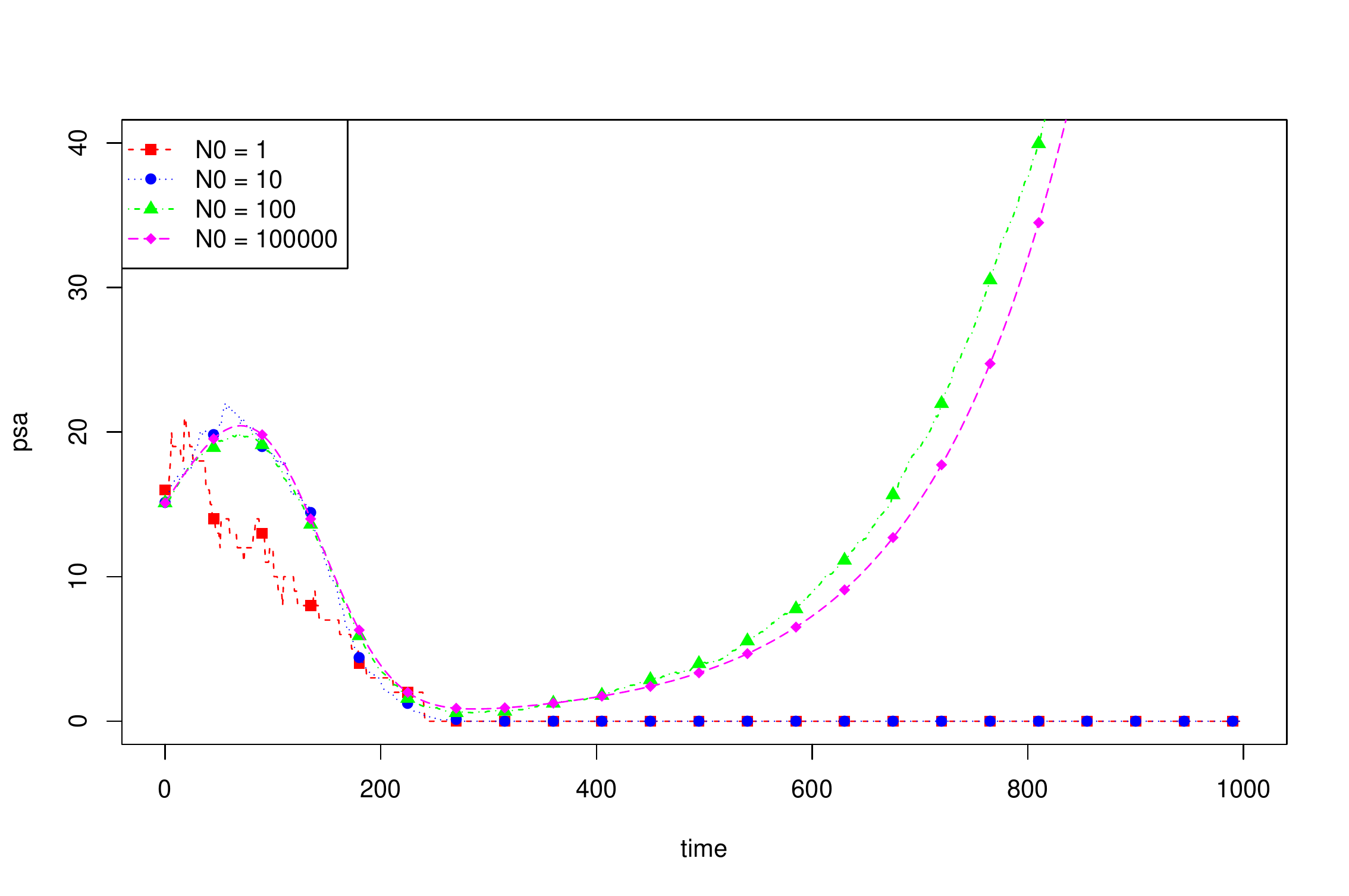} \label{figN0}}
\subfigure[{\footnotesize IAS stochastic, external noise source}]{\includegraphics[width
=0.48\textwidth]{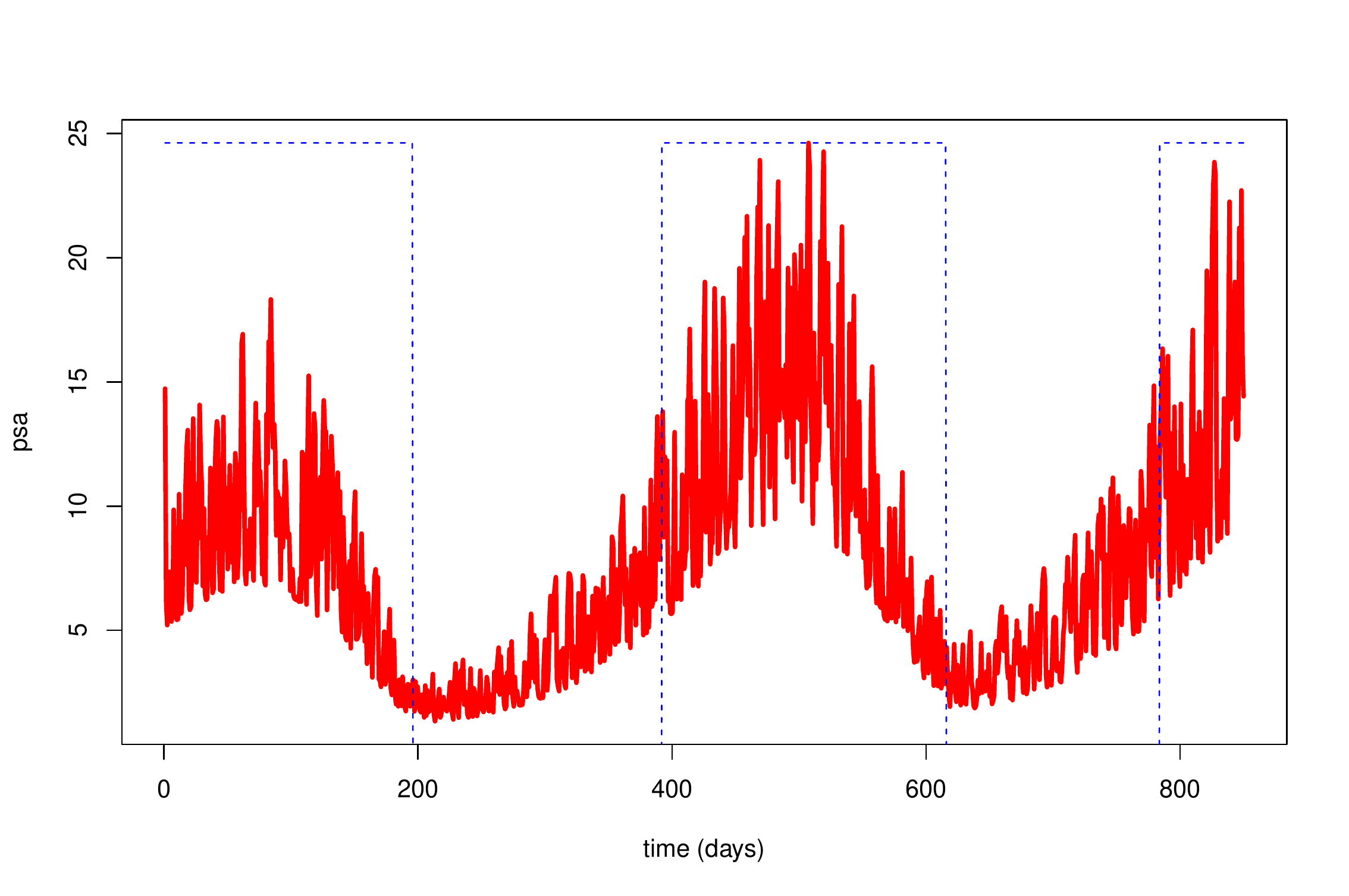} \label{figHidden}}

  \end{center}
  \caption{{\footnotesize (\ref{figCAS} and \ref{figIAS}) Comparison of simulated trajectories of  PSA in the \sCCP-programs
  and in the ODE models, both under CAS and IAS drug dispensation
  policies. Stochastic trajectories are generated by a
  Gillespie-like simulation
  algorithm~\cite{SB:Gillespie:1977:gillespieAlgorithm}. The time unit
  is a day. Parameters of ODE are as in Figure~\ref{tab:ODE}. Scaling parameters
  of the stochastic model are set to $N_0 = 10^5$, $\Omega_Z =
  10^5$, $\Omega_V = 10^5$. PSA of the stochastic model is rescaled
  as $v = \frac{V}{\Omega_V}$ before being plotted, so to have the
  same scale in both modes.
  Initial values of scaled variables are $x = 15$, $y = 0.1$,
  $z=12$. In the CAS policy, chemical castration is always in force.
  The IAS policy adopted is the following: every 4 weeks (28 time
  units) the value of PSA is checked. Drug dispensation is
  interrupted if normalized PSA has dropped below 4, while it is
  resumed when PSA exceeds 10.
  The behaviour of the stochastic model is essentially
  indistinguishable from the behavior of the ODE one. This basically happens for scaling parameters
  greater than $10^3$, hence their precise values do not have a relevant impact in the dynamics, as
  long as they are large enough.
  (\ref{figN0}) Comparison of single trajectories for CAS policy, and different values of $N_0$. 
  (\ref{figHidden}) Single trajectory for IAS policy in the presence of an external random disturbance source.}} \label{figPlots}
\end{figure}

\subsection{Hybrid dynamics and drug dispensation}
\label{sec:events}

The \sCCP-program described above corresponds to the Continuous
Androgen Suppression (CAS)
policy~\cite{SB:Jackson:2004:prostateModelCAS,BIO:CANCER:Brawer:2006:HormonalTherapyProstate}.
As said, more effective drug dispensation policies have been studied
and in particular the Intermittent Androgen Suppression (IAS) has
been proven effective to control prostate tumour
development~\cite{BIO:CANCER:Abrahamsson:2010:IASprostateCancerReview}.
The mathematical rendering of such a policy calls naturally into
play a hybrid model describing the on/off modes of drug
dispensation. In our stochastic program this is done by introducing
a variable $U$ (corresponding to the variable $u$ of the differential equation
model) governing the switch between on/off androgen deprivation
policy. The syntactic feature of $U$ is that it turns out to appear
only in guards and not in rates: it is purely a  control variable.
More specifically, it appears in the new agent corresponding to
androgen hormone production:

{\small
\begin{eqnarray*}
\mathtt{produceANDHc} & \mbox{:-} & [U = 0 \rightarrow Z' = Z +
1]_{\frac{z_0\Omega_Z}{\tau}}.\mathtt{produceANDHc}  +  [U = 1 \rightarrow Z' = Z +
1]_{0}.\mathtt{produceANDHc}
\end{eqnarray*} }
The above agent replaces the corresponding agent
($\mathtt{produceANDH}$) in Figure \ref{tab:sccpProgram}.

Typical drug dispensation policies for androgen deprivation, control
PSA concentration at fixed time intervals (usually every 4 weeks)
and determine whether dispensation should be resumed/suspended by
checking whether specific threshold values are reached. In order to
describe such a policy, we can use \sCCP\ agents with infinite rate
and guards on system time, as described in
Section~\ref{sec:extensionSCCP}. More precisely, we just need a
variable $W$, which will describe in which day the level of PSA will be controlled, initially set to $\delta_t = 28$, and add in parallel to the \sCCP\ program the agent $\mathtt{checkPSA\_on}$, defined by:
{\small
\begin{eqnarray*}
\mathtt{checkPSA\_on} & \mbox{:-} & [Time = W \wedge V < v_{on} \rightarrow W' = W + \delta_t \wedge U' = 0]_\infty.\mathtt{checkPSA\_off}\\
 & + & [Time = W \wedge V \geq v_{on} \rightarrow W' = W + \delta_t]_\infty.\mathtt{checkPSA\_on}\\
\mathtt{checkPSA\_off} & \mbox{:-} & [Time = W \wedge V \geq v_{off} \rightarrow W' = W + \delta_t \wedge U' = 1]_\infty.\mathtt{checkPSA\_on}\\
& + &  [Time = W \wedge V < v_{off} \rightarrow W' = W + \delta_t]_\infty.\mathtt{checkPSA\_off}
\end{eqnarray*} } 
Notice that, according to the semantics of Section \ref{sec:extensionSCCP}, this is a proper hybrid model, as the explicit management of simulation time is a continuous ingredient. However, our semantics machinery in terms of TDSHA allows a much higher degree of flexibility, for instance treating (some of) the system variables as continuous to speed up simulation times.



\section{Experimental results}
\label{sec:results}

In this section we briefly present an experimental analysis of the models of the previous section. More details on the analysis can be found in the supplementary material, available online \cite{My2011COMPMODsupp}. The parameters are fixed to the same values of the ODE model (see Figure~\ref{tab:ODE}). The three scaling factors, $N_0$, $\Omega_Z$ and $\Omega_V$, have been fixed to values that we deemed meaningful and that were checked against real data, see the caption of Figure \ref{figPlots}\footnote{Changing the value of the scaling factors can be seen as assuming a different granularity in counting. For instance, if we have $N$ tumour cells, and we change the reference parameter $N_0$ from 1 to 10, it means that we count how many groups of 10 cells are present in the system.\label{foot:pseudocount}}.

The first analysis that we consider  is the comparison of the
evolution of the agent-based stochastic model with the temporal
evolution of ODE. In Figures \ref{figCAS} and \ref{figIAS} we compare the dynamics of PSA for the CAS and IAS drug dispensation policies, respectively. We show only the value of PSA, as it is the only observable quantity of the model. As we can see, the two models behave essentially in the same way. Even if a similar behaviour had to be expected, the fact that the stochastic system basically shows no noise at all may be seen as quite surprising.
Actually, this phenomenon can be easily explained observing that the variables we are considering are all taking large values, i.e. they correspond to large populations, on the order of millions of cells or millions of molecules (in Figure~\ref{figPlots}, a PSA level of 10 corresponds to 1 million molecules). In these circumstances, the relative magnitude of fluctuations, which is of the order of $\frac{1}{\sqrt{N}}$, is too small to produce significant effects~\cite{SB:Gillespie:2000:ChemicalLangevinEquation}. This essentially means that the variability in behaviour between single cells is lost when we consider large populations: the differences cancel out and the observed behaviour essentially coincides with the average one (cf. also Section 1 of Supplementary Material \cite{My2011COMPMODsupp}).

One of the advantages of having a discrete and stochastic model is that we can also study the behaviour of the model when the
population of cells is small. One way to perform this experiment in this setting is to decrement the parameter $N_0$, with the overall effect of reducing the total number of cancer cells. In
Figure~\ref{figHidden}, we show the evolution of the number of
tumour cells for different values of $N_0$. As expected, as $N_0$
decreases the evolution becomes more noisy. Interestingly, for very small populations  the model changes its behaviour: there is a considerable probability that AI cells can extinguish  before AD cells are eliminated by chemical castration, so that  the tumour tends to be completely eliminated and no relapse can be observed. These low numbers may describe a situation during the initial stages of the tumour. Interestingly, the IAS therapy is less effective for low populations, as it reduces the chance of extinction of AD cells, so that the probability of tumour extinction is considearably lower. Details can be found in Section 2 of the Supplementary Material \cite{My2011COMPMODsupp}.

%

The \sCCP-program model of tumor growth  of Section~\ref{sec:model} essentially does not show any noise in case of large cellular populations. Therefore, we considered possible modifications of the
model to introduce some form of \textit{internal} variability. This approach can be justified in order to check whether the variability in PSA concentration that is observed in real measurements, can be explained by the simple structure of this model. First, we modified the model
adding a random source of variation in the PSA production rate. In Section 3 of the Supplementary Material \cite{My2011COMPMODsupp} we present an experiment in which
the production rate of PSA is no longer a constant with respect to the total
number of tumor cells, but it is variable. In
particular, we assume that production rate of PSA is a random
number, uniformly distributed in the interval
$[0,2\frac{\Omega_V(X+Y)}{N_0}]$. This is accomplished in the
\sCCP-program by introducing a new variable $K$ (the new production rate) and replacing the agent $\mathtt{producePSA}$ with the
following one:
$$\mathtt{producePSA'}\ \mbox{:-}\ [true \rightarrow K' =
Unif(0,2)]_{\infty}.[true \rightarrow V' =
V + 1]_{K\frac{\Omega_V}{N_0}(X+Y)}.\mathtt{producePSA'}.$$ Interestingly, even a large randomness in the PSA production rate does not result in a significant noisy behaviour: also in this case the
fluctuations are averaged out.

We also considered a different modification of the model, in which we try to see whether noise may emerge as a consequence of the interaction with some \textit{external} mechanism, possibly dependent on some global (physiological) condition. Specifically, we added a variable $H$ (for ``hidden'') that can assume just two possible values, namely 0 and 1, representing the presence/absence of an unspecified (physiological) condition. We assume that $H$ switches from 0 to 1 and vice versa twice a day on average, and that the production rate of PSA is subject to a threefold increase when $H=1$. Hence, we added to the model the following agent:
$$\mathtt{hidden}\ \mbox{:-}\ [H = 0\rightarrow H' = 1]_{2} + [H = 1\rightarrow H' = 0]_{2},$$ and we replaced the PSA degradation rate by $(1+2H)$. Trajectories for this model are shown in Figure \ref{figHidden}. 
In this case, we observe noise also in the case of large populations, cf. Section 4 of the Supplementary Material \cite{My2011COMPMODsupp}.

We stress that these last two models do not necessarily have any
biologically significant interpretation. They are just  illustrative examples of possible sources of noise in systems with large populations. 

The analysis we carried out is different from the one of~\cite{SB:Aihara:2010:prostateRoyalSoc}. In this paper, noise is introduced in the model as an additional disturbance term in the ODEs (hence using Stochastic Differential Equations). In this case, however, the model does not provide any possible explanation of noise in terms of  mechanisms intrinsic to the system under study. On the contrary, our study is aimed at better clarifying if the observed noise can be explained in terms of specific mechanisms, i.e. in terms of intrinsic properties of a given model. Notice that some noise will inevitably be introduced by factors external to the system, for instance by measurement errors.

What we understood from the analysis presented here is that the structure of phenomenological tumour cell growth models, like the one considered in this paper, may not be sufficiently rich to contain internal mechanisms for noise generation. If one is
interested in these issues, then more complex models, taking into account more detailed biological mechanisms, should be considered.
These models can also be easily described in our programming
framework. We stress that this kind of analysis is more easily
carried out in a discrete and stochastic setting.

\vspace{-0.2cm}
\section{Conclusion}
\label{sec:conc}
\vspace{-0.2cm}

%
%
%
%
%

The technique presented in this paper  consists in showing how to step from a differential equation model to a ``program'' model, taking the form of a network of interacting agents. The specific programming language we used, allows us also to introduce a stochastic element (internal to the model and) rendered as the \emph{speed} at which any specific interaction takes place. In particular, we considered a few extensions of \sCCP-programs to describe time-driven events and random updates. General \sCCP-programs, in addition, can easily model cell duplication events in the low level model of Section~\ref{sec:background}, by an unrestricted usage of parallel composition and local variable declaration. In general, communication between agents representing single cells can be easily encoded in an asynchronous setting like the one of \sCCP\, using dedicated variables, playing the role of communication channels, or modelling protein-mediated interaction.
Exploiting the programmability of the shared memory (constraint store), one can easily introduce spatial information \cite{SB:Bortolussi:2009:BioLogic} or more complex cell interaction rules.  Hence, \sCCP-programs allow us to model explicitly geometrically qualified interactions or complex competitive dynamics regulating cells growths and deaths. However, a satisfactory definition of the hybrid semantics for this larger class of \sCCP\ programs is still an open issue.

A programming environment like \sCCP\ allows the construction of a ``wizard'' for fast prototyping of (cancer) cell population dynamics. Among other things, this approach should allow us to easily address such basic questions as the effect and nature of noise, parameter dependencies, logical structure of the interactions, etc. More advanced analysis techniques, like statistical model checking \cite{SbZuliani2009BayesianStatisticalMC}, will further enhance the framework. 

We presented here a quantitative analysis on the nature of noise for a differential equation model of prostate cancer, based on the construction of an agent-based version of the model. Specifically, we showed that the phenomenological interactions of this model are not able to explain observed noise in data. We suggested that a more detailed description of interaction and regulation mechanisms involved is needed to better clarify the noise effects. We plan to further investigate this direction, taking also into account spatial organization of the tumour. Our future work will also benefit from a comparison with experimental data.\footnote{This is not so relevant for the work presented here, given that the stochastic and the ODE model are essentially indistinguishable and given that a comparison with experimental data has been carried out in \cite{SB:Aihara:2008:prostateIAS} for the ODE model.}
%
%




\vspace{-0.3cm}
\begin{thebibliography}{10}
\providecommand{\bibitemdeclare}[2]{}
\providecommand{\urlprefix}{Available at }
\providecommand{\url}[1]{\texttt{#1}}
\providecommand{\href}[2]{\texttt{#2}}
\providecommand{\urlalt}[2]{\href{#1}{#2}}
\providecommand{\doi}[1]{doi:\urlalt{http://dx.doi.org/#1}{#1}}
\providecommand{\bibinfo}[2]{#2}

\vspace{-0.2cm}


\bibitemdeclare{misc}{SB:SBMLwebsite}
\bibitem{SB:SBMLwebsite}
\emph{\bibinfo{title}{SBML website}}.
\newblock \bibinfo{note}{\url{http://www.sbml.org}}.



\bibitemdeclare{article}{BIO:CANCER:Abrahamsson:2010:IASprostateCancerReview}
\bibitem{BIO:CANCER:Abrahamsson:2010:IASprostateCancerReview}
\bibinfo{author}{P.A. Abrahamsson} (\bibinfo{year}{2010}):
  \emph{\bibinfo{title}{Potential benefits of intermittent androgen suppression therapy in the treatment of prostate cancer: a systematic review of literature.}}
\newblock {\sl \bibinfo{journal}{Eur Urol}} \bibinfo{volume}{57}, pp.
  \bibinfo{pages}{49--59}, \doi{10.1016/j.eururo.2009.07.049}.

\bibitemdeclare{article}{PA:Bernardo:1998:EMPAtcs}
\bibitem{PA:Bernardo:1998:EMPAtcs}
\bibinfo{author}{M.~Bernardo} \& \bibinfo{author}{R.~Gorrieri}
  (\bibinfo{year}{1998}): \emph{\bibinfo{title}{A tutorial on EMPA: a theory of
  concurrent processes with nondeterminism, priorities, probabilities and
  time}}.
\newblock {\sl \bibinfo{journal}{Theoret. Comput. Sci.}} \bibinfo{volume}{202},
  pp. \bibinfo{pages}{1–--54}, \doi{10.1016/S0304-3975(97)00127-8}.

\bibitemdeclare{misc}{My2011COMPMODsupp}
\bibitem{My2011COMPMODsupp}
  \emph{\bibinfo{title}{Supplementary Material}}.
\newblock
  \bibinfo{note}{\url{http://www.dmi.units.it/\~bortolu/files/COMPMOD2011supp.pdf}}.

\bibitemdeclare{article}{SB:Bortolussi:2008:BiomodelingSCCP:Journal}
\bibitem{SB:Bortolussi:2008:BiomodelingSCCP:Journal}
\bibinfo{author}{L.~Bortolussi} \& \bibinfo{author}{A.~Policriti}
  (\bibinfo{year}{2008}): \emph{\bibinfo{title}{Modeling Biological Systems in
  Concurrent Constraint Programming}}.
\newblock {\sl \bibinfo{journal}{Constraints}}
  \bibinfo{volume}{13}(\bibinfo{number}{1}), \doi{10.1007/s10601-007-9034-8}.

\bibitemdeclare{article}{PA:Bortolussi:2009:SCCPandODEjournal}
\bibitem{PA:Bortolussi:2009:SCCPandODEjournal}
\bibinfo{author}{L.~Bortolussi} \& \bibinfo{author}{A.~Policriti}
  (\bibinfo{year}{2009}): \emph{\bibinfo{title}{Dynamical systems and
  stochastic programming --- from Ordinary Differential Equations and back}}.
\newblock {\sl \bibinfo{journal}{T. Comp. Sys. Bio.}}, \bibinfo{volume}{XI} pp.
  \bibinfo{pages}{216-267}, \doi{10.1007/978-3-642-04186-0\_11}.
 
\bibitemdeclare{inproceedings}{SB:Bortolussi:2009:CompMod}
\bibitem{SB:Bortolussi:2009:CompMod}
\bibinfo{author}{L.~Bortolussi} \& \bibinfo{author}{A.~Policriti}
  (\bibinfo{year}{2009}): \emph{\bibinfo{title}{Hybrid Semantics of Stochastic
  Programs with Dynamic Reconfiguration}}.
\newblock In: {\sl \bibinfo{booktitle}{Proc. of CompMod}}, \doi{10.4204/EPTCS.6.5}.

\bibitemdeclare{inproceedings}{SB:Bortolussi:2009:BioLogic}
\bibitem{SB:Bortolussi:2009:BioLogic}
\bibinfo{author}{L.~Bortolussi} \& \bibinfo{author}{A.~Policriti}
  (\bibinfo{year}{2009}): \emph{\bibinfo{title}{Tales of Spatiality in
  stochastic Concurrent Constraint Programming}}.
\newblock In: {\sl \bibinfo{booktitle}{Proc. of Bio-Logic}}. 

\bibitemdeclare{article}{PA:Bortolussi:2010:HybridDynamicsStochProg:TCS}
\bibitem{PA:Bortolussi:2010:HybridDynamicsStochProg:TCS}
\bibinfo{author}{L.~Bortolussi} \& \bibinfo{author}{A.~Policriti}
  (\bibinfo{year}{2010}): \emph{\bibinfo{title}{Hybrid Dynamics of Stochastic
  Programs}}.
\newblock {\sl \bibinfo{journal}{Theor. Comp. Sc.}} \bibinfo{volume}{411}(\bibinfo{number}{20}), pp. \bibinfo{pages}{2052-2077}, \doi{10.1016/j.tcs.2010.02.008}.



\bibitemdeclare{article}{BIO:CANCER:Brawer:2006:HormonalTherapyProstate}
\bibitem{BIO:CANCER:Brawer:2006:HormonalTherapyProstate}
\bibinfo{author}{M.~K. Brawer} (\bibinfo{year}{2006}):
  \emph{\bibinfo{title}{Hormonal Therapy for Prostate Cancer}}.
\newblock {\sl \bibinfo{journal}{Rev Urol}} \bibinfo{volume}{8}, pp.
  \bibinfo{pages}{S35--S47}. 

\bibitemdeclare{article}{SB:Ciocchetta:2009:BioPEPAevents}
\bibitem{SB:Ciocchetta:2009:BioPEPAevents}
\bibinfo{author}{F.~Ciocchetta} (\bibinfo{year}{2009}):
  \emph{\bibinfo{title}{{Bio-PEPA} with Events}}.
\newblock {\sl \bibinfo{journal}{T. Comp. Sys. Bio.}} \bibinfo{volume}{11},
  pp. \bibinfo{pages}{45--68}, \doi{10.1007/978-3-642-04186-0\_3}.

\bibitemdeclare{inbook}{SB:HillstonCiocchetta:2008:ProcessAlgebraSysBio}
\bibitem{SB:HillstonCiocchetta:2008:ProcessAlgebraSysBio}
\bibinfo{author}{F.~Ciocchetta} \& \bibinfo{author}{J.~Hillston}
  (\bibinfo{year}{2008}): \emph{\bibinfo{title}{Formal methods for
  computational systems biology}}, chapter \bibinfo{chapter}{Process algebras in systems biology}, pp. \bibinfo{pages}{265--312}.
\newblock \bibinfo{publisher}{Springer-Verlag}, \doi{10.1007/978-3-540-68894-5\_8}.

\bibitemdeclare{article}{SB:HillstonCiocchetta:2009:bioPEPA}
\bibitem{SB:HillstonCiocchetta:2009:bioPEPA}
\bibinfo{author}{F.~Ciocchetta} \& \bibinfo{author}{J.~Hillston}
  (\bibinfo{year}{2009}): \emph{\bibinfo{title}{Bio-PEPA: A framework for the
  modelling and analysis of biological systems}}.
\newblock {\sl \bibinfo{journal}{Theor. Comp. Sc.}}
  \bibinfo{volume}{410}(\bibinfo{number}{33-34}), pp. \bibinfo{pages}{3065 --
  3084}, \doi{10.1016/j.tcs.2009.02.037}.

\bibitemdeclare{book}{STOC:Davis:1993:PDMP}
\bibitem{STOC:Davis:1993:PDMP}
\bibinfo{author}{M.H.A. Davis} (\bibinfo{year}{1993}):
  \emph{\bibinfo{title}{Markov Models and Optimization}}.
\newblock \bibinfo{publisher}{Chapman \& Hall}. 

\bibitemdeclare{book}{PA:Balbo:1995:ModelingGSPN}
\bibitem{PA:Balbo:1995:ModelingGSPN}
\bibinfo{author}{M.~Ajmone~Marsan}, \bibinfo{author}{G.~Balbo},
  \bibinfo{author}{G.~Conte}, \bibinfo{author}{S.~Donatelli} \& \bibinfo{author}{G.~Franceschinis}
  (\bibinfo{year}{1995}): \emph{\bibinfo{title}{Modelling with Generalized
  Stochastic Petri Nets}}.
\newblock \bibinfo{publisher}{Wiley}. 

\bibitemdeclare{article}{SB:Gillespie:2000:ChemicalLangevinEquation}
\bibitem{SB:Gillespie:2000:ChemicalLangevinEquation}
\bibinfo{author}{D.~Gillespie} (\bibinfo{year}{2000}):
  \emph{\bibinfo{title}{The chemical Langevin equation}}.
\newblock {\sl \bibinfo{journal}{Journal of Chemical Physics}}
  \bibinfo{volume}{113}(\bibinfo{number}{1}), pp. \bibinfo{pages}{297--306}, \doi{10.1063/1.481811}.

\bibitemdeclare{article}{SB:Gillespie:1977:gillespieAlgorithm}
\bibitem{SB:Gillespie:1977:gillespieAlgorithm}
\bibinfo{author}{D.T. Gillespie} (\bibinfo{year}{1977}):
  \emph{\bibinfo{title}{Exact Stochastic Simulation of Coupled Chemical
  Reactions}}.
\newblock {\sl \bibinfo{journal}{J. of Phys. Chem.}}
  \bibinfo{volume}{81}(\bibinfo{number}{25}), \doi{10.1021/j100540a008}.

\bibitemdeclare{article}{PA:Hermanns:2000:PAforPerformanceEvaluation}
\bibitem{PA:Hermanns:2000:PAforPerformanceEvaluation}
\bibinfo{author}{H. Hermanns}, \bibinfo{author}{U. Herzog} \&
  \bibinfo{author}{J.P. Katoen} (\bibinfo{year}{2002}):
  \emph{\bibinfo{title}{Process algebra for performance evaluation}}.
\newblock {\sl \bibinfo{journal}{Theor. Comp. Sci.}}
  \bibinfo{volume}{274}(\bibinfo{number}{1-2}), pp. \bibinfo{pages}{43--87}, \doi{10.1016/S0304-3975(00)00305-4}.

\bibitemdeclare{article}{SB:Aihara:2008:prostateIAS}
\bibitem{SB:Aihara:2008:prostateIAS}
\bibinfo{author}{A.M. Ideta}, \bibinfo{author}{G.~Tanaka},
  \bibinfo{author}{T.~Takeuchi} \& \bibinfo{author}{K.~Aihara}
  (\bibinfo{year}{2008}): \emph{\bibinfo{title}{A mathematical model of
  intermittent androgen suppression for prostate cancer}}.
\newblock {\sl \bibinfo{journal}{Nonlinear Science}} \bibinfo{volume}{18}, pp.
  \bibinfo{pages}{593--614}, \doi{10.1007/s00332-008-9031-0}.

\bibitemdeclare{article}{SB:Jackson:2004:prostateModelCAS}
\bibitem{SB:Jackson:2004:prostateModelCAS}
\bibinfo{author}{T.~L. Jackson} (\bibinfo{year}{2004}): \emph{\bibinfo{title}{A
  mathematical model of prostate tumor growth and androgen-independent
  relapse}}.
\newblock {\sl \bibinfo{journal}{Disc Cont Dyn Sys B}} \bibinfo{volume}{4}, pp.
  \bibinfo{pages}{187--201}, \doi{10.3934/dcdsb.2004.4.187}.

\bibitemdeclare{inproceedings}{SbZuliani2009BayesianStatisticalMC}
\bibitem{SbZuliani2009BayesianStatisticalMC}
\bibinfo{author}{S.K. Jha}, \bibinfo{author}{E.M. Clarke},
  \bibinfo{author}{C.J. Langmead}, \bibinfo{author}{A.~Legay},
  \bibinfo{author}{A.~Platzer} \& \bibinfo{author}{P.~Zuliani} (\bibinfo{year}{2009}):
  \emph{\bibinfo{title}{A Bayesian Approach to Model Checking Biological
  Systems}}.
\newblock In: {\sl \bibinfo{booktitle}{Proc. of the CMSB}}, pp.
  \bibinfo{pages}{218--234}, \doi{10.1007/978-3-642-03845-7\_15}. 

\bibitemdeclare{inproceedings}{SB:Lecca:2011:TumorShrinkage}
\bibitem{SB:Lecca:2011:TumorShrinkage}
\bibinfo{author}{P.~Lecca}, \bibinfo{author}{O.~Kahramanogullari},
  \bibinfo{author}{D.~Morpurgo}, \bibinfo{author}{C.~Priami} \&
  \bibinfo{author}{R.~Soo} (\bibinfo{year}{2011}):
  \emph{\bibinfo{title}{Modelling the tumor shrinkage pharmacodynamics with
  BlenX}}.
\newblock In: {\sl \bibinfo{booktitle}{Proc. of ICCABS}}, \doi{10.1109/UKSIM.2011.24}.

\bibitemdeclare{article}{SB:Mazza:2009:TumorGrowthP}
\bibitem{SB:Mazza:2009:TumorGrowthP}
\bibinfo{author}{T.~Mazza} \& \bibinfo{author}{M.~Cavaliere}
  (\bibinfo{year}{2009}): \emph{\bibinfo{title}{Cell Cycle and Tumor Growth in
  Membrane Systems with Peripheral Proteins}}.
\newblock {\sl \bibinfo{journal}{Electron. Notes Theor. Comput. Sci.}}
  \bibinfo{volume}{227}, pp. \bibinfo{pages}{127--141}, \doi{10.1016/j.entcs.2008.12.108}.

\bibitemdeclare{inbook}{PA:Mode:2005:SemiMarkovProcesses}
\bibitem{PA:Mode:2005:SemiMarkovProcesses}
\bibinfo{author}{C.J. Mode} (\bibinfo{year}{2005}):
  \emph{\bibinfo{title}{Semi-Markov Processes}}.
\newblock \bibinfo{publisher}{John Wiley \& Sons, Ltd}. 

\bibitemdeclare{book}{STOC:Norris:1997:MarkovChains}
\bibitem{STOC:Norris:1997:MarkovChains}
\bibinfo{author}{J.~R. Norris} (\bibinfo{year}{1997}):
  \emph{\bibinfo{title}{Markov Chains}}.
\newblock \bibinfo{publisher}{Cambridge University Press}. 

\bibitemdeclare{article}{BIO:CANCER:Rao:2008:PSAdiscovery}
\bibitem{BIO:CANCER:Rao:2008:PSAdiscovery}
\bibinfo{author}{A.R. Rao}, \bibinfo{author}{H.G. Motiwala} \&
  \bibinfo{author}{O.M.A. Karim} (\bibinfo{year}{2008}):
  \emph{\bibinfo{title}{The discovery of Prostate-Specific Antigen}}.
\newblock {\sl \bibinfo{journal}{BJU Int.}} \bibinfo{volume}{101}, pp.
  \bibinfo{pages}{5--10}, \doi{10.1111/j.1464-410X.2007.07138.x}.

\bibitemdeclare{article}{StocSkulj2009ImpreciseMC}
\bibitem{StocSkulj2009ImpreciseMC}
\bibinfo{author}{D.~Skulj} (\bibinfo{year}{2009}):
  \emph{\bibinfo{title}{Discrete time Markov chains with interval
  probabilities}}.
\newblock {\sl \bibinfo{journal}{Int. J. Approx. Reasoning}}
  \bibinfo{volume}{50}(\bibinfo{number}{8}), pp. \bibinfo{pages}{1314--1329}, \doi{10.1016/j.ijar.2009.06.007}.

\bibitemdeclare{article}{SB:Aihara:2010:prostateRoyalSoc}
\bibitem{SB:Aihara:2010:prostateRoyalSoc}
\bibinfo{author}{G.~Tanaka}, \bibinfo{author}{Y.~Hirata}, \bibinfo{author}{S.L.
  Goldenberg}, \bibinfo{author}{N.~Bruchovsky} \& \bibinfo{author}{K.~Aihara}
  (\bibinfo{year}{2010}): \emph{\bibinfo{title}{Mathematical modelling of
  prostate cancer growth and its application to hormone therapy}}.
\newblock {\sl \bibinfo{journal}{Phyl Trans Royal Soc A}}
  \bibinfo{volume}{368}, pp. \bibinfo{pages}{5029--5044}, \doi{10.1098/rsta.2010.0221}.

\bibitemdeclare{book}{SB:Wilkinson:2006:StochasticModellingSB}
\bibitem{SB:Wilkinson:2006:StochasticModellingSB}
\bibinfo{author}{D.~J. Wilkinson} (\bibinfo{year}{2006}):
  \emph{\bibinfo{title}{Stochastic Modelling for Systems Biology}}.
\newblock \bibinfo{publisher}{Chapman \& Hall}. 

\end{thebibliography}

\end{document}